\documentclass[11pt]{article}
\usepackage{moriond,epsfig}

\usepackage[T1]{fontenc}
\usepackage{amssymb}
\usepackage[totalwidth=520pt, totalheight=680pt, left=15mm, letterpaper]{geometry}
 
\def\Journal#1#2#3#4{{#1} {\bf #2}, #3 (#4)}

\def\AJ{\em Astron. Jour.}
\def\ApJ{\em Astroph. Jour.}
\def\ApJL{\em Astroph. Jour. Let.}
\def\ApJS{\em Astroph. Jour. Suppl. Ser.}
\def\MNRAS{\em Mon. Not. Roy. Astr. Soc.}
\def\PASA{\em Publ. Astr. Soc. Pac.}

\def\be{\begin{equation}}
\def\ee{\end{equation}}

\DeclareMathAlphabet{\mathitbf}{OML}{cmm}{b}{it}

\def\mrm{\mathrm{m}}
\def\etal{{et~al.}}
\def\bmd{\mathitbf{d}}
\def\bmg{\mathitbf{g}}
\def\bmv{\mathitbf{v}}

\begin{document}
\vspace*{0.7in}
\title{IS THE 2MASS DIPOLE CONVERGENT?}

\author{Micha{\l} CHODOROWSKI \& Maciej BILICKI}
\address{N. Copernicus Astronomical Center,\\Bartycka 18, 00-716 Warsaw, Poland}

\author{Gary A. MAMON}
\address{Institut d'Astrophysique de Paris (UMR 7095: CNRS \& UPMC),\\98 bis Bd Arago, F-75014 Paris, France}

\author{Thomas JARRETT}
\address{Spitzer Science Center, California Institute of Technology,\\Pasadena, CA 91125, USA}

\maketitle\abstracts{We study the growth of the clustering dipole of galaxies from the Two Micron All Sky Survey (2MASS). We find that the dipole does not converge before the completeness limit of the 2MASS Extended Source Catalog, i.e.\ up to about $300\,\mathrm{Mpc}/h$. We compare the observed growth of the dipole with the theoretically expected, conditional growth for the $\Lambda$CDM power spectrum and cosmological parameters constrained by \textit{WMAP}. The observed growth turns out to be within 1$\sigma$ confidence level of the theoretical one, once the proper observational window of the 2MASS flux dipole is included. For a contrast, if the adopted window is a top hat, then the predicted dipole grows significantly faster and converges to its final value at a distance of about $200\,\mathrm{Mpc}/h$. We study the difference between the top-hat window and the window for the flux-limited 2MASS survey and we conclude that the growth of the 2MASS dipole at effective distances greater than $200\,\mathrm{Mpc}/h$ is only apparent. Eventually, since for the window function of 2MASS the predicted growth is consistent with the observed one, we can compare the two to evaluate $\beta\equiv \Omega_\mrm^{0.55}\slash b$. The result is $\beta \simeq 0.38\pm0.05$, which gives 
a rough estimate of $\Omega_\mrm\simeq0.2\pm0.1$.}

\section{Introduction}
For more than 20 years now, many attempts have been made to measure the peculiar gravitational acceleration of the Local Group of galaxies (LG). Such a measurement is possible with the use of all-sky galaxy catalogues, under the assumption that visible (luminous) matter is a good tracer of the underlying density field.

The general procedure is to calculate the so-called \textit{clustering dipole} of a galaxy survey, $\bmd$, and use it to infer the acceleration of the LG, $\bmg$. Within linear theory of gravitational instability, these two quantities are proportional, although under several assumptions. First, the survey should cover the whole sky; second, the observational proxy of the gravitational force should have known properties and last but not least, the survey should be deep enough for the dipole to be convergent to the final value that we want to measure. If one or more of these assumptions are not met, the dipole $\bmd$ is a biased estimator of the acceleration $\bmg$ and the inference of the latter from the former may be done only when the mentioned effects are properly accounted for.

We focus on the third of these effects, i.e. the question of convergence of the clustering dipole, using the data from the Two Micron All Sky Survey (2MASS, Skrutskie \etal\ \cite{Skr}) Extended Source Catalog (XSC). The XSC is complete for sources brighter than $K_s\simeq13.5^\mathrm{mag}$ ($\sim\!2.7\,\mathrm{mJy}$) and resolved diameters larger than $\sim\!10$--$15\,\mathrm{arcsec}$. The near-infrared flux is particularly useful for the purpose of large-scale structure studies as it samples the old stellar population, and hence the bulk of stellar mass, and is minimally affected by dust in the Galactic plane (Jarrett \cite{Jar04}).

The gravitational instability scenario of large-scale structure formation relates peculiar velocities of galaxies with their peculiar gravitational accelerations. In linear theory, this relation has a particularly simple form (Peebles \cite{Pe80}):
\be\label{eq:v and g} \bmv=\beta\,\bmg\;, \ee
where $\beta\equiv \Omega_\mrm^{0.55}\slash b$, $\Omega_\mrm$ is the current value of the cosmological density parameter of non-relativistic matter, $b$ is the linear biasing parameter and the acceleration $\bmg$ is expressed in units of velocity. Applying Eq.~(\ref{eq:v and g}) to the motion of the Local Group as a gravitationally bound system, we can evaluate the $\beta$ parameter from LG's peculiar velocity and acceleration. If we have some additional knowledge on the biasing, we can estimate the value of the density parameter $\Omega_\mrm$.

The velocity of the LG is known from the dipole component of the temperature distribution of cosmic microwave background (CMB). When reduced to the barycenter of the LG, it equals to $v_\mathrm{CMB}=622\pm35\,\mathrm{km\slash s}$ in the direction $(l,b)=(272^\circ\pm3^\circ\!,\,28^\circ\pm5^\circ)$ in Galactic coordinates. As for the acceleration of the LG, it can be estimated from a two-dimensional all-sky catalog, i.e. one containing astro- and photometric data only, provided that we know the behaviour of the mass-to-light ratio in the band(s) of the survey.

\section{Growth of the 2MASS dipole}
\begin{figure}
\begin{center}
\psfig{figure=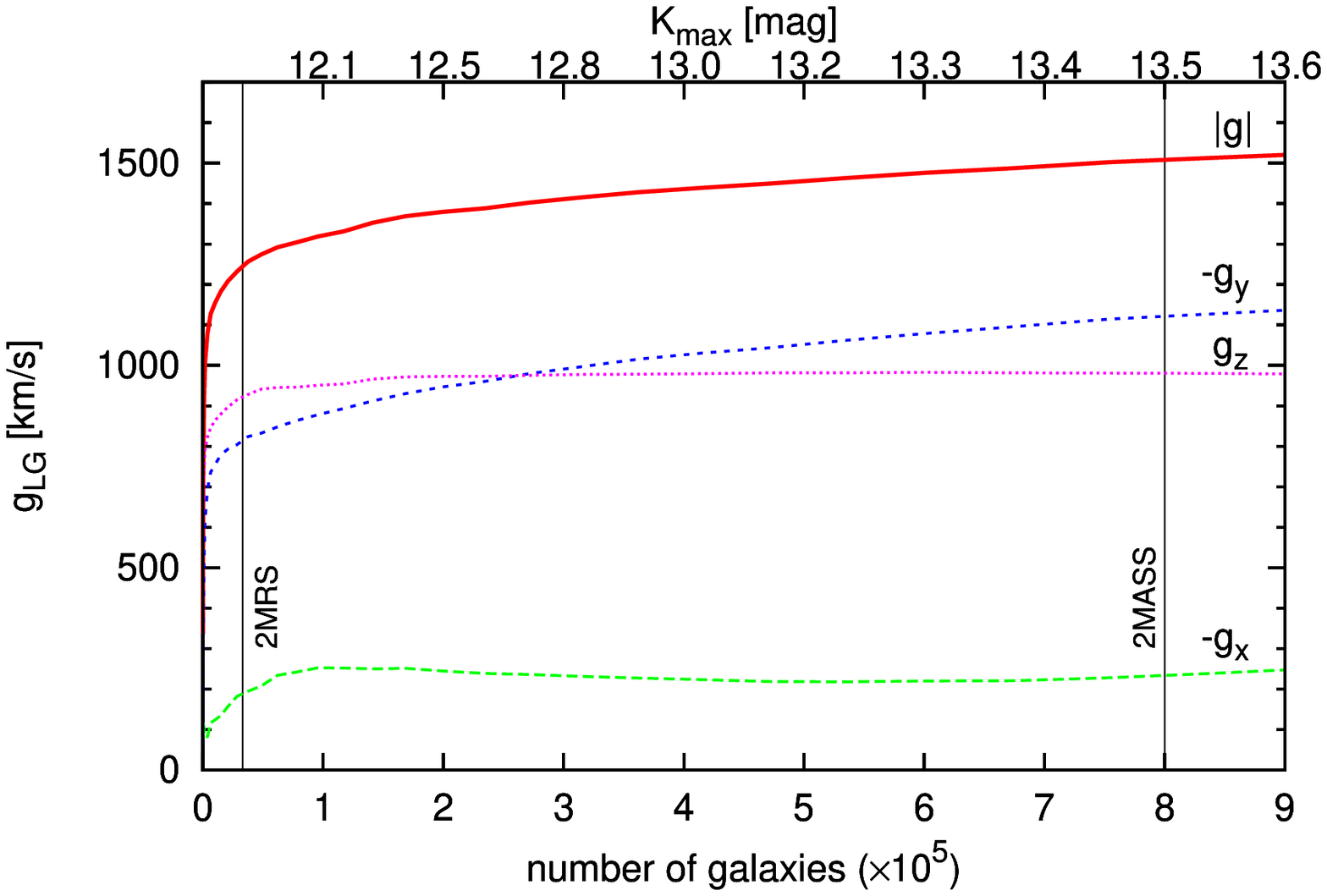,angle=-90,width=263pt}
\psfig{figure=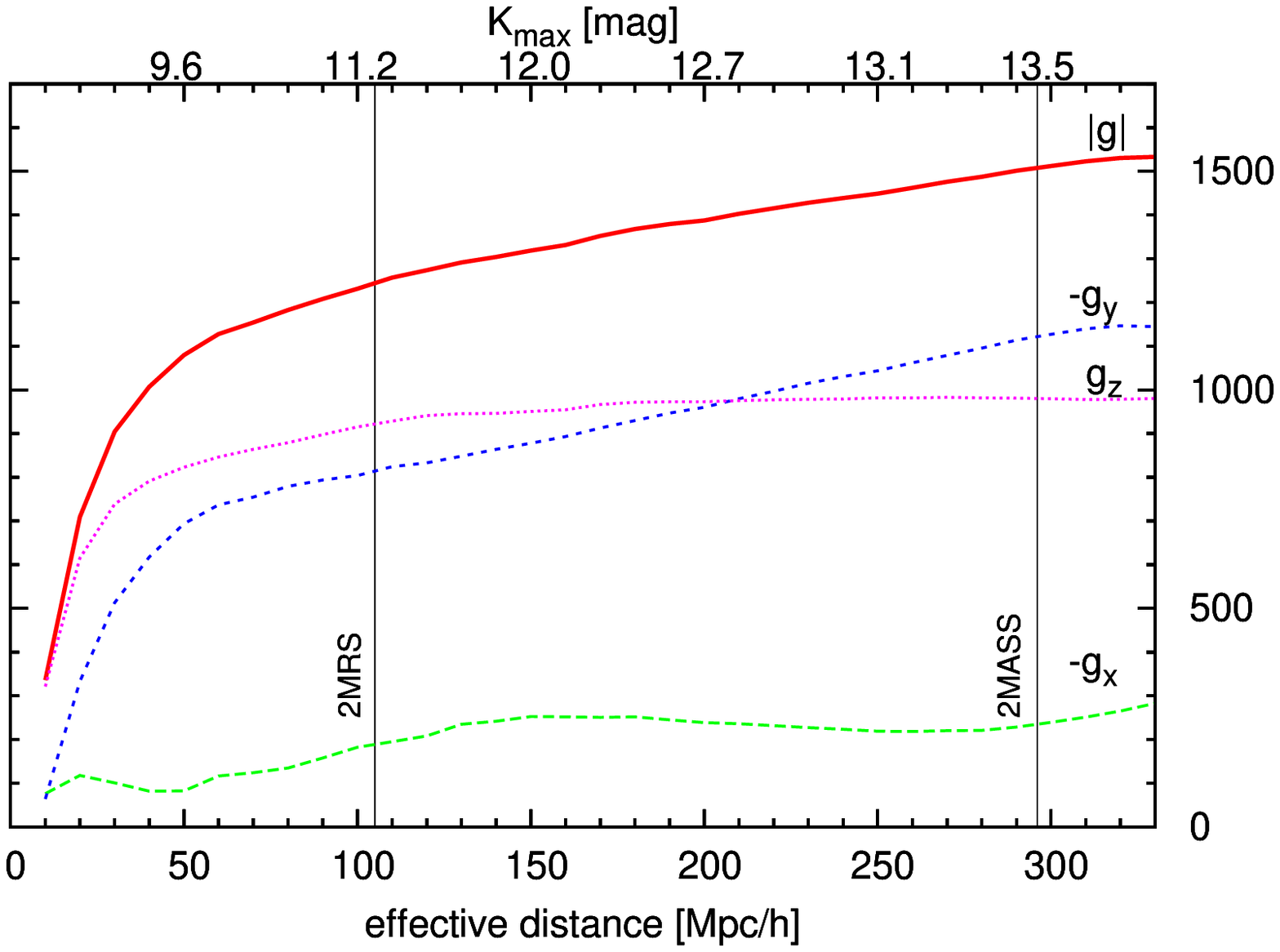,angle=-90,width=251pt}
\end{center}
\caption{Growth of the 2MASS clustering dipole. \textit{Left}: as a function of the number of galaxies used for the calculation. \textit{Right}: as a function of the effective distance. Top axes show corresponding cut-off magnitudes. Solid lines represent the amplitude of the dipole; dotted and dashed lines illustrate Galactic Cartesian components.\label{fig:growth}}
\end{figure}
The main interest of the study presented here was to examine the growth of the clustering dipole of galaxies from 2MASS XSC as a function of increasing depth of the sample. For that purpose we used positions and fluxes in the near-infrared $K_s$ band ($2.16\,\mu\mrm$), obtained from the catalog~\footnote{\texttt{http://pegasus.phast.umass.edu/data\_products/all\_sky\_catalog/index.html}}, corrected for extinction and other effects, such as Zone of Avoidance. The growth was calculated by decrementing the minimum flux of the objects in the sample (i.e.\ incrementing the maximum $K_s$ magnitude). Results are presented in Fig.~\ref{fig:growth}. The left panel shows the growth of the dipole as a function of the number of galaxies, together with Galactic Cartesian components. Such a presentation was used by Maller \etal\ (2003) \cite{Mal03} and would suggest convergence of the dipole, just as they concluded. However, a linear scale in the number of galaxies on the $x$-axis is not a convenient measure of the sample depth. What is needed is a linear scale \textit{in distance} on the abscissa. For that purpose we related magnitudes of galaxies with effective distances. As an estimator of the latter we propose the median value of distance given the flux. Its calculation involves the luminosity function (LF) parameters for a given band. For the LF in the $K_s$ band as given by Jones \etal\ \cite{HJ}, we find $r_\mathrm{eff} \simeq 0.59 \times 10^{\,0.2\,K}\,\mathrm{Mpc}/h$ for the magnitude $K$. This proxy of distance is used in the right panel of Fig.\ \ref{fig:growth}. The growth of the clustering dipole up to the completeness limit of 2MASS XSC ($\approx\! 300\,\mathrm{Mpc}/h$) is now evident.

\section{Observations vs. theory}
\begin{figure}
\begin{center}
\psfig{figure=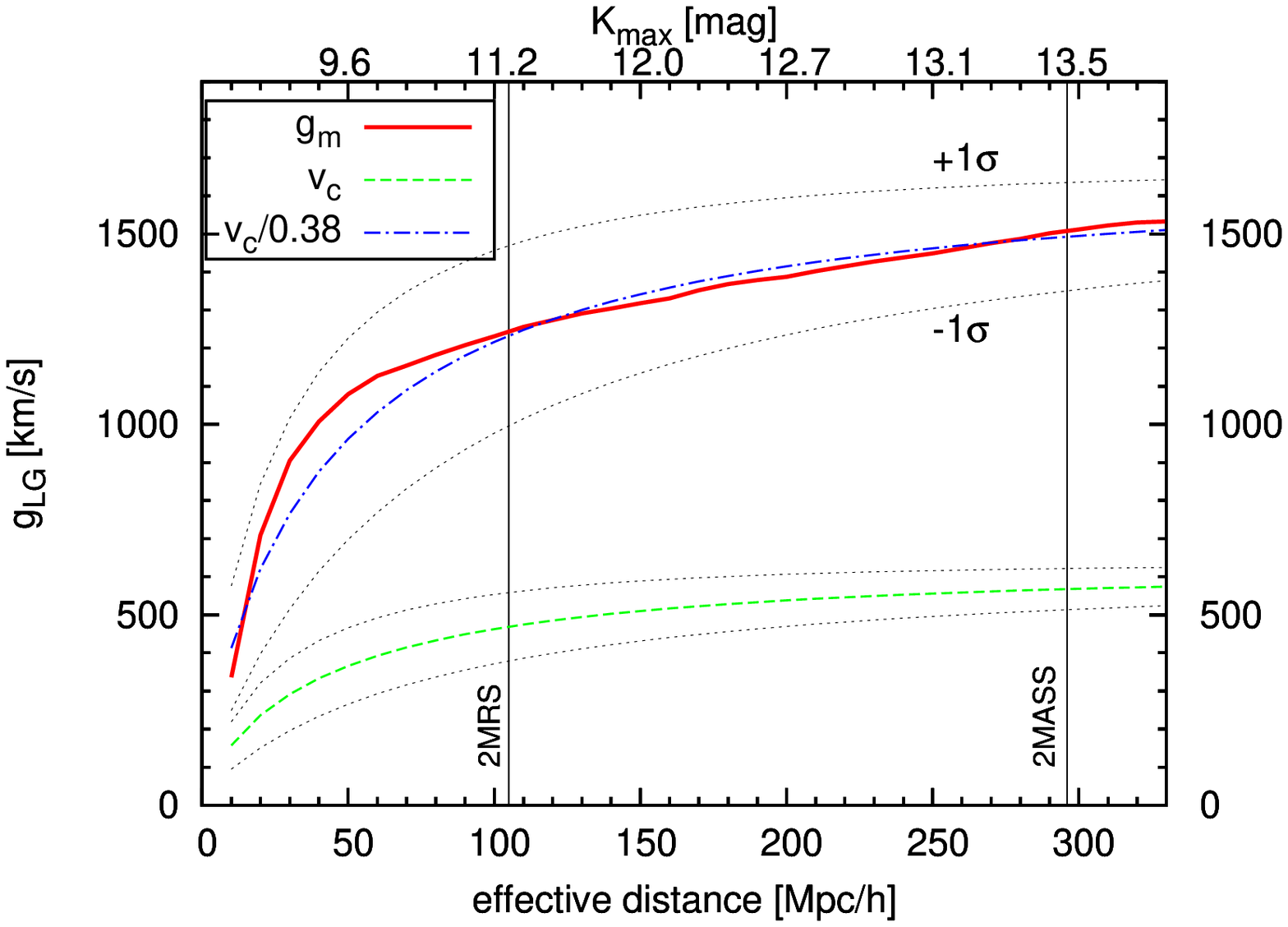,angle=-90,width=258pt}
\psfig{figure=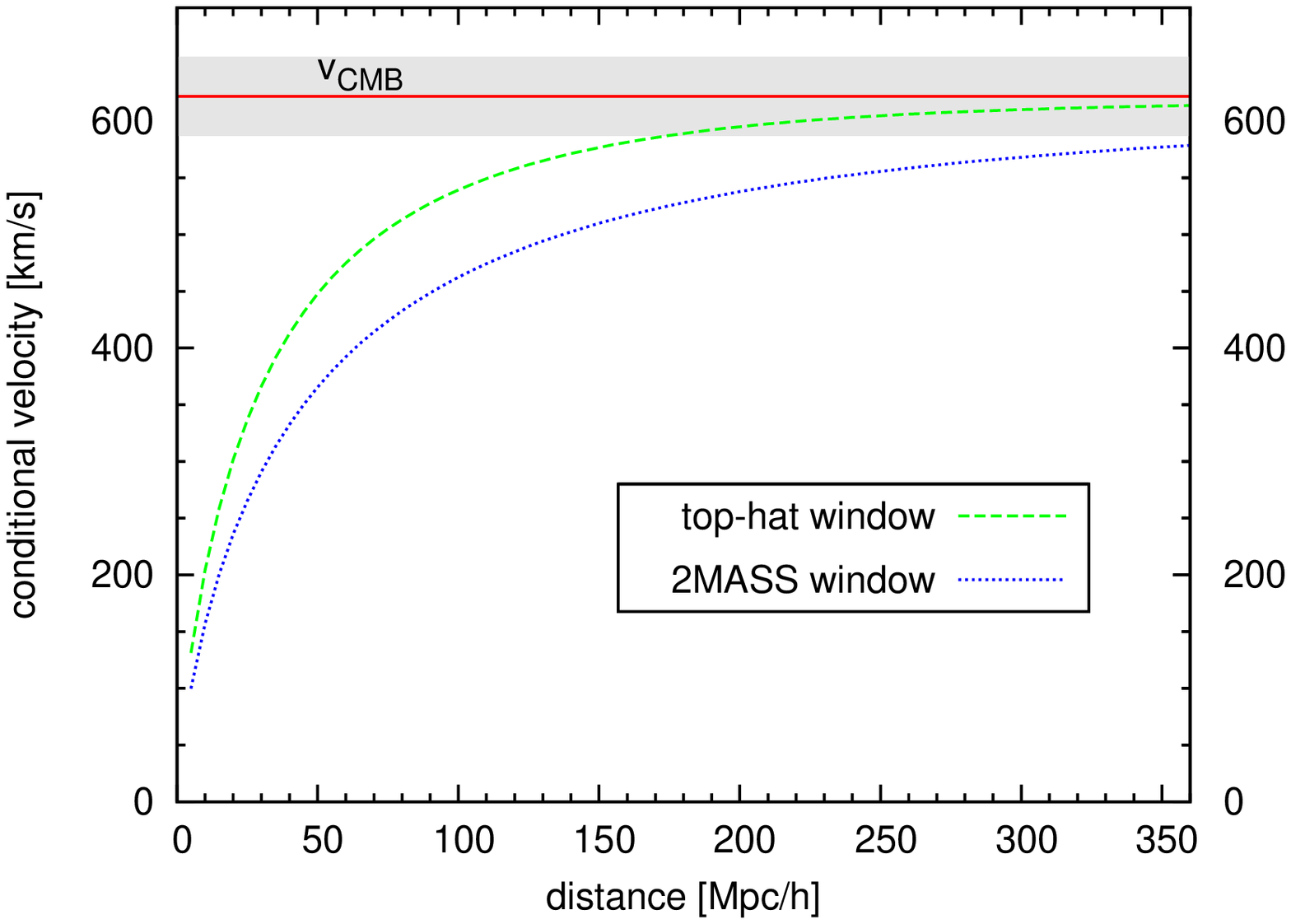,angle=-90,width=258pt}
\end{center}
\caption{\textit{Left}: Growth of the 2MASS clustering dipole (solid line), compared with the theoretical expectation for conditional velocity of the Local Group in the $\Lambda$CDM model (dashed line) and the latter rescaled by the $\beta$ parameter (dot-dashed line). The dotted lines show 1$\sigma$ variances for the theoretical curves. 
\textit{Right}: Theoretically predicted conditional velocity of the Local Group for known $v_\mathrm{CMB}$ using two different observational windows: the flux window of 2MASS (dotted line) and top{}-{}hat (dashed line). The horizontal solid line is the velocity of the Local Group with respect to the CMB and the shaded strip represents confidence intervals of the $v_\mathrm{CMB}$ measurement.
\label{fig:compar}}
\end{figure}
Taken at face value, this divergence is consistent with the findings of some other authors, who used various datasets and methods. On the other hand, it contradicts existing claims of convergence (even at scales as small as $60$--$100\,\mathrm{Mpc}/h$). However, this discrepancy most probably stems from the different nature of catalogs and methods used for the calculation, and in particular is due to distinct \textit{observational windows}. Knowledge of these windows is necessary to correctly confront such results and is also essential if we want to make comparisons with theoretical expectations.

We thus pose the following question: is the behaviour of the 2MASS flux dipole consistent with the predictions of the currently accepted cosmological model, namely Lambda--Cold-Dark-Matter ($\Lambda$CDM)? Looking for the answer, we compare our results with the expectation value for the amplitude of the acceleration of the LG \textit{knowing} its peculiar velocity. The relevant formulae can be found in Juszkiewicz \etal\ \cite{JVW} and \mbox{Lahav \etal\ \cite{LKH}} Results are presented in Fig.~\ref{fig:compar} (left panel). The expectation value of the acceleration (dot-dashed line) is obtained from the conditional velocity (dashed line) rescaled by the $\beta$ parameter according to Eq.~(\ref{eq:v and g}). This velocity, calculated from eq.\ (8a) of Juszkiewicz \etal\ \cite{JVW}, includes the CDM power spectrum (PS) of density fluctuations and the observational window of the 2MASS flux-limited survey. For the former we use the linear PS of CDM as given by Eisenstein \& Hu \cite{EH98} together with \textit{WMAP} 5-year cosmological parameters (Hinshaw \etal\ \cite{Hinsh}). The latter is the window calculated by Chodorowski \etal\ \cite{CCBCC}

As can be seen in the left panel of Fig. \ref{fig:compar}, the
observed growth of the 2MASS dipole is well within the 1$\sigma$ range
of the theoretical prediction rescaled by $\beta=0.38$. In the right
panel of Fig.~\ref{fig:compar} we compare the prediction for the 2MASS
window with the one for top-hat (i.e.\ a sphere of radius $R$). The
latter window is appropriate for all-sky catalogues that include
redshifts (as for example the 2MASS Redshift Survey, 2MRS,
\mbox{Huchra \etal\ \cite{Huch}).} It is noticeable that the
expectation value of the conditional velocity (without rescaling) for
the 2MASS window is far from converging to the limit of
$v_\mathrm{CMB}=622\,\mathrm{km\slash s}$ even for $r$ approaching
$300\,\mathrm{Mpc}/h$. On the contrary, the theoretically expected
dipole measured through a top-hat window has clearly converged to
$v_\mathrm{CMB}$ for $r\simeq300\,\mathrm{Mpc}/h$. It should be noted,
however, that even for all-sky redshift surveys, such as 2MRS, the
convergence of the dipole is rather unlikely before some
$200\,\mathrm{Mpc}/h$, contrarily to the measurement of Erdo\u{g}du
\etal\ \cite{Erdogdu}, where it is claimed that the contribution from
structure beyond $6000\,\mathrm{km}/\mathrm{s}$
$(=60\,\mathrm{Mpc}/h)$ is negligible (but cf.\ Lavaux
\etal\ \cite{Lav10}). We can see that for the latter distance, the
conditional velocity for the top-hat window has reached only 75\% of
its final value.

The slower convergence of the dipole measured through the 2MASS window compared to the top-hat case is easy to understand. In redshift space, the contribution to the flux dipole from, for instance, a galaxy cluster will come from distances of plus/minus several megaparsecs around the true value (the so-called \textit{Fingers of God}). In the case of angular data only, numerous faint galaxies of the cluster will contribute to the flux at effective distances that are {\em significantly\/} greater than the true distance of the cluster. 

The preliminary errorbars for our measurement of $\beta$ can be evaluated from the error in $v_\mathrm{CMB}$ ($\sim\!6\%$) and taking the theoretical $\sigma_g$ as a measure of the uncertainty on $g$ ($\sim\!9\%$). This altogether results in $\Delta\beta\simeq0.05$. The value of $\beta=0.38\pm0.05$ is in accordance with Erdo\u{g}du \etal\ \cite{Erdogdu} and only slightly below the lower confidence limit of Pike \& Hudson \cite{PH05}. Using the biasing $b_{K_s}=1.1\pm0.2$ from Maller \etal\ (2005) \cite{Mal05} we obtain a rough estimate of the density parameter: $\Omega_\mrm\simeq0.2\pm0.1$.

More details of the study presented here will be given in a forthcoming paper. In the future, we also plan to apply the maximum likelihood method to the 2MASS data, as described in Chodorowski \etal\ \cite{CCBCC}, for an \textit{optimal} calculation of the $\beta$ parameter and its errors.

\section*{Acknowledgments}
This publication makes use of data products from the Two Micron All Sky Survey, which is a joint project of the University of Massachusetts and the Infrared Processing and Analysis Center/California Institute of Technology, funded by the National Aeronautics and Space Administration and the National Science Foundation. M.B. and M.C. were partially supported by the Polish Ministry of Science and Higher Education under grant N N203 0253 33, allocated for the period 2007--2010.

\end{document}